\documentclass{mn2e}
\usepackage{epsf}

\def\msun{{\rm M_{\odot}}}

\title[Fuelling Active Galactic Nuclei]
{Fuelling Active Galactic Nuclei}

\author[A. R. King and J. E. Pringle]
{A. R. King$^1$ and
J.E. Pringle$^{1,2}$\\ $^1$Theoretical Astrophysics Group, University
of Leicester, Leicester LE1 7RH\\ $^2$Institute of Astronomy,
University of Cambridge, Madingley Road, Cambridge CB3 0HA}

\date{\today}

\volume{000}

\pagerange{\pageref{firstpage}--\pageref{lastpage}} \pubyear{2006}

\begin{document}

\label{firstpage}

\maketitle

\begin{abstract}

We suggest that most nearby active galactic nuclei are fed by a series
of small--scale, randomly--oriented accretion events. Outside a
certain radius these events promote rapid star formation, while within
it they fuel the supermassive black hole. We show that the events have
a characteristic time evolution. This picture agrees with several
observational facts. The expected luminosity function is broadly in
agreement with that observed for moderate--mass black holes. The spin
of the black hole is low, and aligns with the inner disc in each
individual feeding event. This implies radio jets aligned with the
axis of the obscuring torus, and uncorrelated with the large--scale
structure of the host galaxy. The ring of young stars observed about
the Galactic Centre are close to where our picture predicts that star
formation should occur.

\end{abstract}


\section{Introduction}

It is now generally accepted that the nuclei of most galaxies contain
supermassive black holes. Assembling the observed hole masses by
accretion at plausible efficiency accounts for the emission of active
galactic nuclei (AGN) over cosmic time (Soltan, 1982; Yu \& Tremaine,
2002). This implies that essentially every galaxy is intermittently
active, in phases when its black hole grows. Only a small fraction of
galaxies are observed to be active (e.g. Heckman et al., 2004), so we
know that these phases are short.

Yet we still do not understand how gas gets down to the black
hole. There are a number of candidates, including galaxy mergers
(major and minor), bars, bars within bars, turbulence in the ISM,
stellar mass loss, and viscous accretion discs. However, the nature of
the problem is often underestimated. For example, in a recent review
article, Wada (2004) discusses `Fuelling Gas to the Central Region of
Galaxies' in terms of bringing gas inwards to around 100 pc from the
central black hole. But, as emphasised by Shlosman, Begelman \& Frank
(1990), the gas must get to remarkably small radii before the inner
viscous accretion disc can take over and bring the gas to the hole
within the observed activity timescale. The inflow timescale through a
disc of size 1~pc is already $\sim 10^9$ yr, even with maximal
assumptions about viscosity. For shorter AGN phases, or less efficient
viscosity, the gas must be fed to the central disc at still smaller
radii.

Observations show a wide range of activity, from highly luminous
quasars at redshifts up to $z \simeq 6$, through the barely detectable
nuclei in LINERS, to the weak activity of our own Galactic Centre. It
seems likely that all of the fuelling processes discussed so far may
play a role in at least some galaxies. In this paper we focus
attention on the fuelling of the nucleus in a normal Seyfert galaxy,
with a central black hole of mass $M \sim 10^7 - 10^8$ M$_\odot$,
(Ferrarese et al., 2001; Heckman et al., 2004; Denney et al., 2006)
and luminosity $L \sim 0.1 L_E$ (with $L_E$ the Eddington luminosity)
thus accreting at a rate $\dot{M} \sim 0.2 \msun$~yr$^{-1}$. In a
typical Seyfert event lasting, say, $10^6$~yr, this would imply an
accreted mass $\Delta M \sim 2 \times 10^5$ M$_\odot$ (e.g. Emsellem,
2002, 2004; Martini, 2004). This is ridiculously small compared with
the gas available in a typical spiral galaxy, and we immediately see
that we are not concerned with major gas flows involving spiral arms
or galactic bars. Feeding the black hole at such modest rates needs
only some small--scale event to drop a little material of low angular
momentum into the central regions of the galaxy.

We have two pieces of evidence suggesting what such an event might
be. First, Kinney et al. (2000; see also Nagar \& Wilson, 1999) found
that in a sample of nearby ($z \le 0.031$) Seyfert galaxies, the
direction of the jet from the central black hole, and therefore
presumably the orientation of the central regions of the accretion
disc, were unrelated to the orientation of the disc of the host
galaxy. Second, Schmitt et al (2003) surveyed extended [O III]
emission in a sample of nearby Seyfert galaxies, and found that
although the [O III] emission is well aligned with the radio, there is
no correlation between its orientation and the major axis of the host
galaxy. Assuming that the orientation of the [O III] emission is
governed by the geometry of the inner torus, of typical radius 0.1 --
1.0 pc (e.g. Antonucci, 1993), this means that the central disc flow
has angular momentum unrelated to that of most of the gas in the host
galaxy.

This tells us two things (see also the discussion in Kinney et al.,
2000). First, in line with our inference above, if the galactic disc
supplied the torus material via bar--driven disc evolution, or
grand--design nuclear spirals, or similar mechanisms, some means of
randomising its rotation axis in the central regions is needed.  For
without some such mechanism we expect such processes to leave evidence
of the angular momentum of the galactic disc, which is not seen. There
is currently no suggestion as to what this mechanism might be,
although one can appeal to the fact that the galactic disc is much
thicker than the radial scale of a few parsecs that we are interested
in. In the absence of a randomising mechanism the gas must come from
outside the galaxy as part of a succession of (very) minor
mergers. Where in a galaxy a very small merging satellite deposits its
gas is not straightforward to compute (Velasquez \& White, 1999;
Taylor \& Babul, 2001). Kendall, Magorrian \& Pringle (2003) find that
if such small merging satellites are to deposit gas near the nucleus
then their initial orbits must be fairly accurate shots. Kendall et
al. (2003) also conclude that such small merging satellites which do
manage to reach the nucleus arrive there on more or less randomly
oriented orbits.

Second, the inner disc (specifying the radio jet axis) remains aligned
with the torus. In an earlier paper (King \& Pringle, 2006, Section 3)
we showed that repeated small--scale fuelling has precisely this
effect. It causes counteralignment of the hole (King et al., 2005) in
about one--half of all accretion events, and thus spindown, which is
more efficient than spinup.  Once the hole's mass has doubled, its
angular momentum is always smaller in magnitude than that of any
accretion event. It then always aligns its spin with the inner disc
and thus the torus.

Thus we have argued that the fuelling of low luminosity AGN proceeds
via a series of randomly oriented, small--scale accretion events,
acting directly in the region of the central black hole. The material
deposited in such an event is likely to settle quickly into a ring or
disc of material within a few local dynamical (or orbital) timescales
$t_{\rm dyn}$, where
\begin{equation}
t_{\rm dyn} = 2.9 \times 10^2 \left( \frac{M}{10^8 M_\odot} \right)^{-1/2}
\left( \frac{R}{0.1 {\rm pc}} \right)^{3/2} \, {\rm yr}.
\end{equation}
It is well known that such a disc is likely to be self--gravitating
outside some radius $R_{\rm sg} \sim 0.01 - 0.1$~pc (Shlosman et al.,
1990; Collin--Souffrin \& Dumont, 1990, Hur\'{e} et al.,
1994). Further, in the outer regions of these discs the cooling
timescales are sufficiently short that self--gravity is likely to
cause star formation, rather than enhanced angular momentum transport
(Shlosman \& Begelman, 1989; Collin \& Zahn, 1999). For the discs we
consider in this paper, which are relatively thin (because of
efficient cooling) and of low mass ($M_{\rm disc} \ll M_{\rm hole})$
self--gravity appears first in modes with azimuthal wavenumber $m \approx
R/H$, producing transient spiral waves which transport angular
momentum (Anthony \& Carlberg, 1988; Lodato \& Rice, 2004, 2005). It
is reasonable to suppose that where the disc is locally
gravitationally unstable in this way, most of the gas forms stars.

We therefore propose the hypothesis that all, or at least most of, the
gas initially at radii $R > R_{\rm sg}$ is either turned into stars,
or expelled by those stars which do form, on a rapid (almost
dynamical) timescale (see also Shlosman \& Begelman, 1989). This
corresponds to a nuclear starburst of the kind often associated with
AGN (e.g. Scoville 2002). A starburst like this lasts around $3 \times
10^6$~yr in terms of its ionising flux (O and early B stars). We
suppose further that all the gas which is initially at radii $R <
R_{\rm sg}$ forms a standard accretion disc, which slowly drains on to
the black hole and powers the AGN. We stress that under the simple
form of this hypothesis the nuclear starburst and the AGN are two
different manifestations of the same accretion event, but do not feed
one another. We note that such a picture provides at least a
qualitative explanation for the existence of the ring(s) of young
stars seen around the black hole in the centre of the Milky Way
(Genzel et al., 2003).

In Section 2, we investigate the properties of the accretion disc at
radii $R < R_{\rm sg}$. In Section~3, we consider the time-dependence
of such an event and in Section~4 present a simplified luminosity
function under the assumption that all events are identical, but
randomly timed.

\section{Properties of the disc}

We use the steady disc properties derived by Collin--Souffrin
\& Dumont (1990) in the context of AGN, which are essentially the same
as those derived by Shakura \& Sunyaev (1973) for steady discs in the
context of X--ray binaries. The disc surface density is given as

\begin{eqnarray}
\label{surfdensity}
\lefteqn{\Sigma = 7.5 \times 10^6\left( \frac{\alpha}{0.03} \right)^{-4/5}
\left(\frac{\epsilon}{0.1}\right)^{-3/5}}\nonumber \\
\lefteqn{\times \left( \frac{L}{0.1 L_E}
\right)^{3/5} M_8^{1/5} \left( \frac{R}{R_s} \right)^{-3/5} \,{\rm g
\, cm}^{-2}.}
\end{eqnarray}
Here $\alpha$ is the standard Shakura \& Sunyaev (1973) viscosity
parameter, $\epsilon$ is the accretion efficiency, so that the
luminosity $L$ and accetion rate $\dot{M}$ are related by
\begin{equation}
L = \epsilon \dot{M} c^2,
\end{equation}
also
\begin{equation}
L_E = 1.4 \times 10^{46} M_8 \, {\rm erg \, s}^{-1},
\end{equation}
is the Eddington luminosity, $M_8$ is the black hole mass, $M$, in units of
$10^8$ M$_\odot$, $R$ is the radius and
\begin{equation}
R_s = 2.96 \times 10^{13} M_8 \, {\rm cm}
\end{equation}
is the Schwarzschild radius of the central black hole.

Then the mass of the disc $M(<R)$ interior to radius $R$ is given by
\begin{eqnarray}
\lefteqn{M(<R) = 2.94 \times 10^{34}\left( \frac{\alpha}{0.03} \right)^{-4/5}
\left(\frac{\epsilon}{0.1}\right)^{-3/5}} \nonumber \\
\lefteqn{\times \left( \frac{L}{0.1 L_E}
\right)^{3/5} M_8^{11/5} \left( \frac{R}{R_s} \right)^{7/5} \, {\rm g}.}
\end{eqnarray}

The disc semi--thickness $H$ is given by
\begin{eqnarray}
\lefteqn{\frac{H}{R} = 1.94 \times 10^{-3} \left( \frac{\alpha}{0.03}
\right)^{-1/10} \left(\frac{\epsilon}{0.1}\right)^{-1/5}} \nonumber \\
\lefteqn{\times\left(\frac{L}{0.1 L_E} \right)^{1/5} M_8^{-1/10} 
\left( \frac{R}{R_s}\right)^{1/20}.}
\label{thickness}
\end{eqnarray}

The condition for the disc to become self--gravitating is approximately
(e.g. Pringle, 1981)
\begin{equation}
\frac{M(<R)}{M} \ge \frac{H}{R}.
\label{sg}
\end{equation}
This occurs at radii $R \ge R_{\rm sg}$, where
\begin{eqnarray}
\label{gravradius}
\lefteqn{\frac{R_{\rm sg}}{R_s} = 1.13 \times 10^3 \left( \frac{\alpha}{0.03}
\right)^{14/27} \left(\frac{\epsilon}{0.1}\right)^{8/27}} \nonumber \\
\lefteqn{\times\left(
\frac{L}{0.1 L_E} \right)^{-8/27} M_8^{-26/27}}.
\end{eqnarray}

We note that this implies
\begin{equation}
\label{gravradius2}
{R_{\rm sg} = 0.01 \left( \frac{\alpha}{0.03} \right)^{14/27}
\left(\frac{\epsilon}{0.1}\right)^{8/27}}\left(\frac{L}{0.1 L_E}
\right)^{-8/27} M_8^{1/27}~{\rm pc}
\end{equation}
almost independently of the black hole mass. This arises because
(\ref{gravradius2}) is an integrated form of the standard
steady--state disc equation $\dot M = 3\pi\nu\Sigma$ combined with
(\ref{sg}), which we can express as 
\begin{equation}
R_{\rm sg} \simeq {3\over 2}{H\over R}\alpha c_s{M\over \dot M}
\label{rsg}
\end{equation}
where $c_s$ is a mean sound speed and we note from (\ref{thickness})
that $H/R \simeq$ constant. Encouragingly, we see that $R_{\rm sg}$ is
only slightly smaller than the inner edge $R \sim 0.03$~pc of the ring
of young stars seen around the black hole in the centre of the Milky
Way (Genzel et al., 2003). This is to be expected, as the disc within
$R_{\rm sg}$ must pass its angular momentum to the self--gravitating
region further out which in our picture produces these stars.

The mass inside the radius $R_{\rm sg}$ is given by
\begin{eqnarray}
\label{gravmass}
\lefteqn{M_{\rm sg} = 
2.76 \times 10^5 \left( \frac{\alpha}{0.03} \right)^{-2/27}
\left(\frac{\epsilon}{0.1}\right)^{-5/27}} \nonumber \\
\lefteqn{\times\left( \frac{L}{0.1 L_E}
\right)^{5/27} M_8^{23/27} \, {\rm M}_\odot,}
\label{msg}
\end{eqnarray}
which is of course just $(H/R)M$ with $H/R$ evaluated at $R_{\rm sg}$
(cf. equation~\ref{thickness}).

The accretion rate is given by $\dot{M} = L/\epsilon c^2$, which
implies 
\begin{equation}
\dot{M} = 0.245 
\left(\frac{\epsilon}{0.1}\right)^{-1} \left( \frac{L}{0.1 L_E}
\right) M_8 \, {\rm M}_\odot \, {\rm y}^{-1}.
\end{equation}
Then the evolution timescale of the disc is given by $\tau_{\rm sg} =
M_{\rm sg}/\dot{M}$, which gives
\begin{eqnarray}
\label{gravtime}
\lefteqn{\tau_{\rm sg} 
= 1.12 \times 10^6 \left( \frac{\alpha}{0.03} \right)^{-2/27}
\left(\frac{\epsilon}{0.1}\right)^{22/27}} \nonumber \\ 
\lefteqn{\times\left( \frac{L}{0.1 L_E}
\right)^{-22/27} M_8^{-4/27} \,  {\rm y}}.
\end{eqnarray} 
Again we note that $\tau_{\rm sg} \sim (H/R)(M/\dot M)$.

\section{Properties of an accretion event}

In the previous Section we established the general properties of the
non--self--gravitating disc, assuming that it was accreting steadily,
and giving rise to a luminosity $L$. Of course this is not precisely
what we require for the present problem. What we are assuming is that
the disc, of initial mass $M_{\rm sg}$, is deposited by the accretion event
in some unknown configuration, and then evolves due to viscosity. As
for most accretion discs it is fair to assume that initially most of
the mass, and most of the angular momentum of the disc, is
predominantly at large radius, and therefore at around $R_{\rm sg}$. But
whatever the initial configuration, within a timescale of at most
$\tau_{\rm sg}$ given by eq.(\ref{gravtime}) the disc evolves to resemble a
steady disc at radii $R \le R_{\rm sg}$ (e.g. Pringle, 1981). Thus using the
formulae derived in Section~2 gives a good approximation to the actual
disc properties after a brief initial period. We can now use this to
estimate the properties of the disc, and therefore of the AGN event,
as it evolves.

From eqn.(\ref{surfdensity}) we see that $\Sigma \propto
{\dot{M}}^{3/5} R^{-3/5}$. For a steady disc we know (Pringle 1981)
that $\dot{M} \propto \nu \Sigma$, where $\nu$ is the viscosity. From
these two relationships we can deduce that for these discs $\nu
\propto \Sigma^{2/3} R$. For such discs, similarity solutions (Pringle
1991) imply that at late times the accretion rate, and hence
luminosity, varies with time as $L \propto t^{-19/16}$. Thus we expect
luminosity evolution roughly to follow
\begin{equation}
\label{Levolution}
L = L_{\rm init} [1 + (t/\tau_{\rm sg})]^{-19/16}.
\end{equation}

\section{Implications for the luminosity function}

Heckman et al. (2004) find that most of the current accretion on to
black holes is on to those with masses in the range $10^7 - 10^8$
M$_\odot$. In addition they find that of low mass black holes, with
masses $M < 3 \times 10^7$ M$_\odot$, only 0.2 per cent are growing at
a rate which accounts for 50 per cent of the fuelling. They point out
that this implies that strong fuelling therefore only lasts for a time
$t_{\rm fuel} \sim 0.002 \times t_H$, where $t_H \sim 1.4 \times
10^{10}$ years is the age of the universe. Thus the total time during
which strong fuelling is taking place is around $t_{\rm fuel} \sim 3
\times 10^7$ years, or, if there are $N$ fuelling events per black
hole per Hubble time, it implies that each event lasts around
$t_{\rm ev} \sim 3 \times 10^7/N$ yr (Heckman et al., 2004).
\begin{figure}
\centerline{\epsfxsize9cm \epsfbox{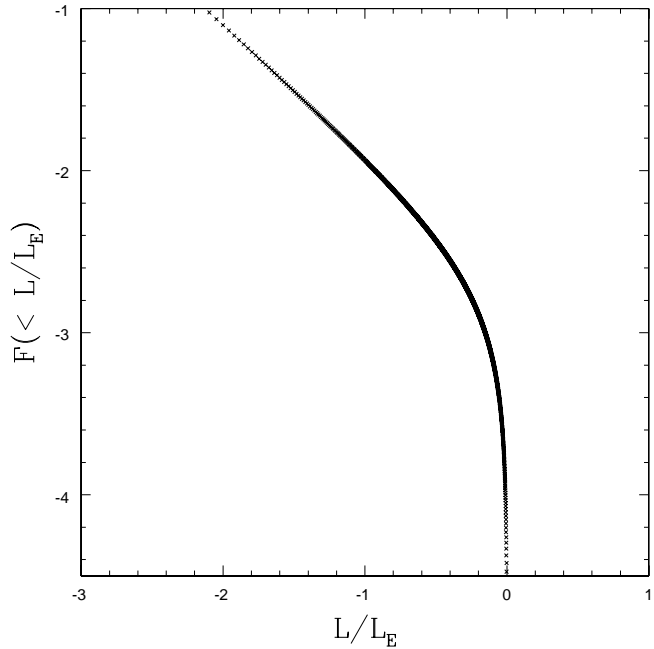}}
\caption{Form of the AGN luminosity function predicted by the fuelling
process, and subsequent disc evolution, discussed in this paper. The
fraction $F$, of those sources with luminosities less than $L/L_E$ is
shown as a function of $L/L_E$. This is similar in form to those
presented in Figure 3 of Heckman et al., 2004}
\label{fuelfig}
\end{figure}

Accordingly we make some simplifying assumptions to work out the
luminosity function we might expect from our model. We take $\alpha =
0.03$ and $\epsilon = 0.1$, and consider black holes with mass
$10^7M_{\odot}$ as representative of the range making the major
contribution to the distribution of accreting black holes. We assume
that these black holes undergo random, but identical, accretion events
such that the initial luminosity of each event is $L_E$. In this case,
we see from the above that the initial disc mass is $M_{\rm sg} \sim 5.95
\times 10^4$ M$_\odot$, and the initial evolution timescale is $\tau_{\rm sg}
\sim 2.41 \times 10^5$ yr. Equating this timescale to the length of
each event, $t_{\rm ev}$ we see that the number of events has to be
about $N =0.002t_H/\tau_{\rm sg} \sim 116$, and therefore the average time
between events is $t_{\rm rep} \sim t_H/N = \tau_{\rm sg}/0.002 \sim 1.2
\times 10^8$ yr.
 
Writing $f = L/L_E$, our assumption implies that the initial value of
$f$ is 1. The average final value is $f_{\rm end} = f_{\rm
in}/ (1 + \lambda_{\rm end}^{19/16})$, where $\lambda_{\rm end} =
t_{\rm rep}/\tau_{\rm sg} = 500$, and so the maximum possible range
of $f$ is $(1 + \lambda_{\rm end})^{19/16}$. Of course observations
cannot probe this full range, and from Figure 3 of Heckman
et al. (2004) we see that for $10^7$ M$_\odot$ black holes the
observed range of $f$ is around 40 (corresponding to $\lambda \ge
18$ and $N \le t_H/18 \tau_{\rm sg} = 540$).

Inverting equation~\ref{Levolution} we find that
\begin{equation}
\frac{t}{\tau_{\rm sg}} = f^{-16/19}  - 1,
\end{equation}
for $0 \le t \le \tau_{\rm sg}$. Then assuming that fuelling events for
different black holes are independent, we find that the fraction
$F(>f)$ of black holes with luminosities $> f$ is given by
\begin{equation}
F(>f) = \frac{f^{-16/19} - 1}{f_{\rm end}^{-16/19} - 1}.
\end{equation}
We plot this in Figure 1. As can be seen, this curve has a similar
form to the distributions of low--mass black holes ($3\times
10^6M_{\odot} - 3\times 10^7M_{\odot}$) in Figure 3 (Left) of Heckman et
al. (2004) in the observed range $1 \la f \la 40$. Given the
simplicity of our assumptions (identical independent fuelling events
etc) we regard this as encouraging.

\section{Conclusions}

We have suggested that the feeding of most nearby active galactic
nuclei proceeds via a series of small--scale, randomly--oriented
accretion events, rather than large--scale events bearing the imprint
of the host galaxy. Outside a certain radius these events cause rapid
star formation, while within it they feed the supermassive black
hole. This picture implies a characteristic time decay of each event
and implies a luminosity function broadly in agreement with that
observed for moderate--mass black holes. The chaotic nature of the
accretion keeps the black hole spin low, allowing each individual
feeding event to produce radio jets aligned with the axis of the
obscuring torus, which itself is uncorrelated with the large--scale
structure of the host galaxy. Our picture predicts that star formation
should occur at radii comparable with those of the ring of young stars
observed about the Galactic Centre. In an earlier paper (King \&
Pringle, 2006) we showed that small--scale feeding events allow
supermassive black holes to reach large ($\ga 10^9{\rm M}_{\odot}$)
masses at high redshifts $\sim 6$. We conclude that small--scale
chaotic feeding offers a promising explanation of the growth of most
supermassive black holes, and thus for the activity of galactic
nuclei.

\section{Acknowledgments} 

ARK acknowledges a Royal Society--Wolfson Research Merit Award.
We thank Sergei Nayakshin for stimulating discussions, and the
referee, Isaac Shlosman, for a very helpful report.

\label{lastpage}

\end{document}